\documentclass[a4paper,twoside]{article}

\usepackage{epsfig}
\usepackage{subcaption}
\usepackage{calc}
\usepackage{amssymb}
\usepackage{amstext}
\usepackage{amsmath}
\usepackage{amsthm}
\usepackage{multicol}
\usepackage{pslatex}
\usepackage{apalike}

\usepackage{hyperref}
\usepackage{amsmath}
\usepackage{algorithmic}

\usepackage{LomasCliffStyle}

\begin{document}

\title{Exploring Narrative Economics: An Agent-Based-Modeling Platform that Integrates Automated Traders with Opinion Dynamics}

\author{\authorname{Kenneth Lomas\footnote{ORCID: 0000-0003-2279-1782} and Dave Cliff\footnote{ORCID:0000-0003-3822-9364}} 
\affiliation{Department of Computer Science, University of Bristol, Bristol BS8 1UB, UK}
\email{kl16942@alumni.bristol.ac.uk, csdtc@bristol.ac.uk}
}


\abstract{In seeking to explain aspects of real-world economies that defy easy understanding when analysed via conventional means, Nobel Laureate Robert Shiller has since 2017 introduced and developed the idea of {\em Narrative Economics}, where observable economic factors such as the dynamics of prices in asset markets are explained largely as a consequence of the {\em narratives}\/ (i.e., the stories) heard, told, and believed by participants in those markets. Shiller argues that otherwise irrational and difficult-to-explain behaviors, such as investors participating in highly volatile cryptocurrency markets, are best explained and understood in narrative terms: people invest because they {\em believe}, because they have a heartfelt {\em opinion}, about the future prospects of the asset, and they tell to themselves and others stories ({\em narratives}) about those beliefs and opinions. In this paper we describe what is, to the best of our knowledge, the first ever agent-based modelling platform that allows for the study of issues in narrative economics. We have created this by integrating and synthesizing research in two previously separate fields: {\em opinion dynamics} (OD), and {\em agent-based computational economics} (ACE) in the form of minimally-intelligent trader-agents operating in accurately modelled financial markets. We show here for the first time how long-established models in OD and in ACE can be brought together to enable the experimental study of issues in narrative economics, and we present initial results from our system. The program-code for our simulation platform has been released as freely-available open-source software on {\em GitHub}, to enable other researchers to replicate and extend our work.
~\\
\vspace*{4em}
~\\
To be presented at the 13th International Conference on Agents and  Artificial Intelligence (ICAART2021), Vienna, 4th--6th February 2021.   
}

\onecolumn \maketitle \normalsize \setcounter{footnote}{0} \vfill

\section{\uppercase{Introduction}}
\label{sec:introduction}

In his influential 2017 paper \cite{narrative_econ_paper}, later expanded into the successful 2019 book {\em Narrative Economics: How Stories Go Viral and Drive Major Economic Events} \cite{narrative_econ_book}, Nobel Laureate Robert Shiller introduced the concept of {\em narrative economics} as an overlooked factor in understanding market trends. In brief, Shiller argues that in many markets the movement and maintenance of prices are driven to a significant extent by the stories -- i.e., the narratives -- that market participants tell each other. Shiller draws comparisons between the spread of narratives and the transmission of infectious diseases, and argues that financial bubbles and crashes (most notably in cryptocurrency markets) can plausibly be accounted for as primarily driven by the narratives that traders tell each other, even when those narratives make little sense to outside observers. 

The narratives told in and about a market are externalisations, verbalizations, of the participants' interior beliefs or opinions. In this paper, we present the first results from a novel synthesis of two previously separate fields that both rely on agent-based modelling: our work combines practices from minimal-intelligence {\em agent-based computational economics} (ACE) with ideas developed separately in the research field known as {\em opinion dynamics}. We show here for the first time how existing well-known and widely-used ACE models of trader-agents can be extended so that each trader also holds its own independent opinion, which is our minimal approximation model of Shiller's notion that real traders are influenced by the narratives that they hear, read, and tell. In our work, an individual trader's opinion may be influenced to varying degrees by the opinions of other traders that it interacts with; and the trader's own opinion also directly influences its individual trading activity, i.e. the sequence of bids and/or offers that it quotes into a single central financial exchange that all traders in our model interact with. Our model financial exchange is technically a {\em continuous double auction} (CDA) market operating with a {\em limit order book} (LOB), which is exactly the structure of existing financial markets such as the New York Stock Exchange, and all other major national and international financial exchanges. 

In keeping with the spirit of minimalism that motivates much ACE work, We show here for the first time how zero-intelligence (ZI) and minimal-intelligence (MI) trader-agents can be extended so that each trader also holds its own independent opinion. For consistency with prior work in opinion dynamics (OD) research, we model each trader's opinion as a signed scalar real value, e.g. as a number in the continuous range $[-1.0, +1.0]$: this approach is long-established in OD research, a field that over its multi-decade history has seen developed a succession of models introduced to explore and/or account for observable patterns of opinion dynamics in human societies. In our work we have explored the integration of ZI/MI traders with the following previously-established OD models: the {\em Bounded Confidence} model \cite{krause,hegselmannandkrause}; the {\em Relative Agreement} model \cite{deffuant2002,reexaminingRA}; and the {\em Relative Disagreement} model \cite{RD}. We refer to these three opinion dynamics models as the BC, RA, and RD models respectively. 

The trader-agents that we extend by addition of these OD models are Gode \& Sunder's (1993) {\em Zero Intelligence Constrained} (ZIC), and the {\em Near-Zero-Intelligence} (NZI) trader agents of \cite{Duffy} which minimally extend Gode \& Sunder's ZI approach in such a way that markets populated by NZI traders can exhibit asset-price bubbles. We refer to the extended agent designs as {\em opinionated agents}: we name our opinionated version of ZIC as OZIC, and our opinionated version of NZI as ONZI. For both OZIC and ONZI agents, the bounds of the probability distribution used to randomly generate a trader's bid or offer prices is dependent at least in part on the current value of that agent's opinion-variable; and that opinion variable can change over time as a consequence of interactions with other traders in the market, thereby modelling Shiller's notion of narrative economics: in our system opinions can drive prices, and prices can alter opinions. To the best of our knowledge, we are the first authors to report on such a system, a synthesis of opinion dynamics and market-trading agents, and so the primary contribution of this paper is the modelling platform that we describe for the first time here. The source-code for our system has been placed in the public domain as a freely-available open-source release on {\em GitHub}.\footnote{ 
{\tt github.com/ken-neth/opinion\_dynamics\_BSE.git}}

We evaluate and test the performance of these trading agents, contrasting and comparing the BC, RA, and RD opinion dynamics models, using as our financial-market simulator {\em BSE}, a long-established open-source simulator of a LOB-based financial exchange for a single asset, and freely available in the public domain since 2012 \cite{BSE}. 
This paper summarises \cite{myThesis}, which contains extensive further visualization and discussion of additional results that are not included here. 

In Section~\ref{sec:background} we summarise relevant prior academic literature. Section~\ref{sec:NZItraders} describes near-zero-intelligence traders in more depth. Section~\ref{sec:opinionatedtraders} then introduces our innovation, the addition of opinions to trading-agent models, giving {\em opinionated traders}, and results from simulation studies  running on our platform are presented in Section~\ref{sec:results}.

\section{\uppercase{Background}}
\label{sec:background}

\subsection{Opinion Dynamics}

People are complicated. In particular, how ideas are formed and conveyed to others are difficult to model as there are numerous factors that could affect the behaviour of individuals. Nevertheless we can say, with some degree of certainty, that people hold opinions and these opinions are changed by interacting with the world. Taking this a step further, people communicate and at some point during or after the communication their opinions may alter as a consequence. Given a sufficiently large population we can design models for how their opinions will change over time, i.e. models of the system's opinion dynamics (OD).  
Of course these models make clear assumptions and may not fully encapsulate the inner workings of a person but can nevertheless be useful in understanding problems relying on the opinions of large populations.

One early OD model is given in \cite{deGroot}.
In this model, a group of experts have different opinions on a subject and want to reach a consensus. The experts decide on a format of structured debate where each individual expert has a turn to express their opinion, taking the form of a real number, and at the end every expert updates their own individual opinion, using a fixed weight. The experts continue to take turns sharing their opinions until a consensus is reached. \cite{deGroot} proves that they will always reach a consensus given positive weights. 

A number of later works have analysed the DeGroot model. In \cite{chatterjee} the DeGroot model's treatment of the consensus problem is related to the ergodicity problem in probability theory, which concerns stochastic state spaces where from a given state all possible states are reachable and hence backwards traversal of the state space is difficult. 

The DeGroot model was subsequently analysed by \cite{friedkin}, who described experiments to understand how the model's mean opinions change over time. Choice-shifts are shown by the difference between the final group mean opinion and their initial mean opinion. These experiments showed how individuals in the population could have greater influence on the overall consensus, and Friedkin argued that choice shifts are an inherent problem in discussions of issues where influence is not balanced.

\subsubsection{Bounded Confidence}

A variation on the DeGroot model was described in \cite{krause} and named the {\em Bounded Confidence} (BC) model. In this, all agents in a fixed-size population hold an opinion that is represented as a real number. The agents share their opinions and only update their opinions if they are closer than a given deviation threshold. The reasoning for this is that humans are less likely to have their opinions swayed by someone whose opinion heavily deviates from their own. A formal specification of the BC model is given in \cite{hegselmannandkrause} and summarised as follows: 
given a population of size $n$, $x_i(t)$ represents the opinion of expert $i$ at time $t$. This is updated by:
\[ x_i(t+1) = a_{i1}x_{1}(t) + a_{i2}x_{2}(t) + ... + a_{in}x_{n}(t), \]
where $a_{ij}$ is the confidence factor between experts $i$ and $j$. Crucially the confidence factor between two experts can be zero if the difference in their opinions are too great. Since at each time step opinions change, it is possible that at a much later time step two agents that initially held too-distant opinions can come to be within a sufficiently close range to start to agree.

At the beginning of a simulation, all opinions should be distributed over $[-1, +1]\subset{\cal R}$, with any individuals holding opinions less than or greater than a certain extreme value parameter regarded as \textit{extremists}. As time progresses, experts whose opinions deviate by less than the deviation threshold move closer together according to a confidence factor. The opinions of the experts will converge until the simulation reaches a stable state with do further changes.

\subsubsection{Relative Agreement}

Another well-known Opinion Dynamics model, the {\em Relative Agreement} (RA) model was proposed by  \cite{deffuant2000}. In the RA model experts hold opinions, that are each represented as a real number, but with the difference that they also hold an {\em uncertainty}, which acts like a range around their opinion. The experts communicate and provided the overlap of their uncertainties exceeds the expert's individual uncertainty then they update their opinion and uncertainty by a weight parameter and a Relative Agreement value. 

\begin{figure}[!h]
    \centering
    \includegraphics[width=0.4\linewidth]{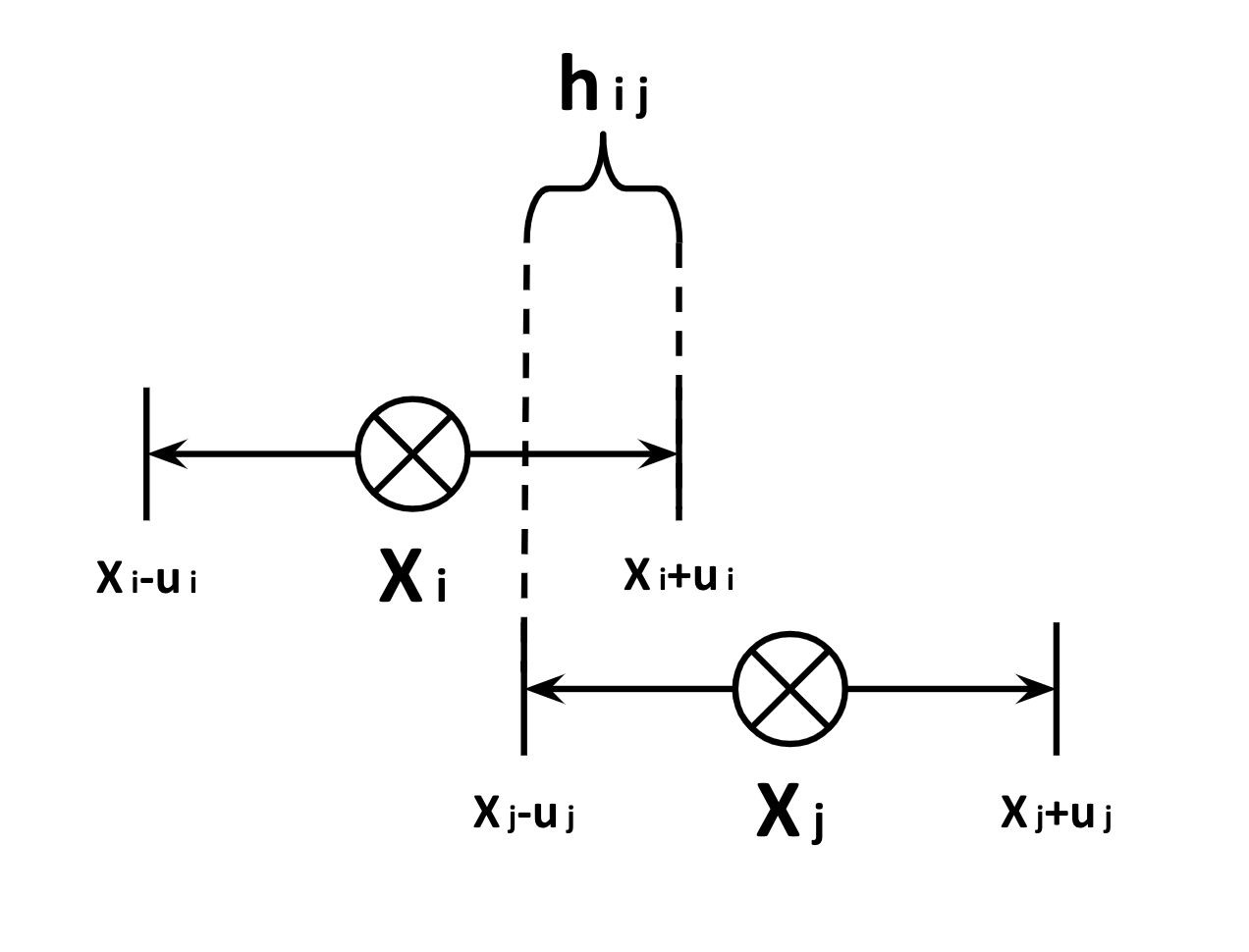}
    \caption{Overlap $h_{ij}$ for experts $i$ and $j$ with opinions $X_i$ and $X_j$ and uncertainties $u_i$ and $u_j$ respectively}
    \label{fig:relativeagreement}
\end{figure}

According to the RA model definition in the Deffuant \textit{et al.} 2000 paper, opinions are updated as follows:
a pair of experts $i$ and $j$ are chosen at random from the population of experts. Firstly, calculate the overlap $h_{ij}$, as illustrated in Figure \ref{fig:relativeagreement},
\[h_{ij} = min (x_i+u_i,x_j+u_j) - max(x_i-u_i,x_j-u_j),\]
where $x_i$ is the real number representation of the opinion of expert $i$, and $u_i$ is the uncertainty of expert $i$ in their own opinion. Then, subtract the size of the non-overlapping part $2u_i- h_{ij}$ so the total agreement of the two experts is given by: 
\[h_{ij}- (2u_i-h_{ij}) = 2(h_{ij}-u_i),\]
and so the RA between $i$ and $j$ is given by:
\[RA_{ij} = 2(h_{ij}-u_i) / 2u_i= (h_{ij}/u_i) - 1\]
Then if $h_{ij} > u_i$, the update is given by:
\[x_j:=x_j+\mu RA_{ij}(x_i - x_j)\]
\[u_j:=u_j+\mu RA_{ij}(u_i - u_j)\]
where $\mu$ is a constant parameter for convergence, similar to the confidence factor in the BC model.
\cite{deffuant2000} show that the RA model converges to an average of $n = w/2u$ opinions as opposed to the BC model that converges to $n = {\rm floor}(w/2u)$ opinions. 

{\em Extremists} were added by  \cite{deffuant2002}, which also describes three modes of convergence that occur with the RA model: central convergence; bipolar convergence; and single-extreme convergence. As with BC, at the beginning of an RA simulation all opinions are randomly distributed over $[-1, +1]\subset{\cal R}$. Central convergence appears as all of the opinions converge towards a stable single central value, around zero. In the case where the opinions converge towards two separate values and reach a stable state, we have bipolar convergence. When all opinions converge towards an extreme value and reach a stable state, exceeding a given extreme parameter, we have single-extreme convergence. 
In a later paper \cite{deffuant2006}, an \textit{asymmetric influence rule} is described where agents that are more convinced of their own opinion exert greater influence upon others.

In \cite{deffuant2002} a metric is used to measure the influence of extremists in a population called the \textit{y metric}. The y metric, or indicator, is given by the formula:
\[y = p_{+}^{2} + p_{-}^{2},\]
where $p_{+}$ denotes the proportion of experts that were initially moderate but held a positive extreme opinion by the end of the simulation, and $p_{-}$ denotes the proportion of experts that were initially moderate but held a negative extreme opinion by the end of the simulation. Deffuant \textit{et al.} use the $y$ metric as an indicator of convergence type, i.e.\ central convergence at $y=0$, bipolar convergence at $y=0.5$, and single extreme convergence at $y=1$.

\subsubsection{Relative Disagreement}

The RA model has been shown to successfully simulate useful convergences in populations with extremists initialized. A more recent model, introduced in \cite{RD}, and  called the Relative Disagreement (RD) model improves on the RA model by introducing probability $\lambda$ of an update occurring and the idea of \textit{reactance}. In  \cite{RD} the RD model was shown to achieve the same opinion convergences as the RA model without the need for initialising the population with extremists.

Reactance is the motivation to disagree with an opinion. In psychology it has been rationalised as a desire to exercise freedom when that freedom is under threat \cite{reactance}. It is an important part of how people behave and how they come to hold certain opinions. The RD model incorporates the idea of reactance by having individuals' opinions diverge when they disagree to enough of a degree. In contrast to $h_{ij}$ in RA, $g_{ij}$ is the non overlapping distance calculated by:
\[g_{ij} = max(x_i - u_i, x_j - u_j) - min(x_i + u_i, x_j + u_j)\]

\begin{figure}[!h]
    \centering
    \includegraphics[width=0.4\linewidth]{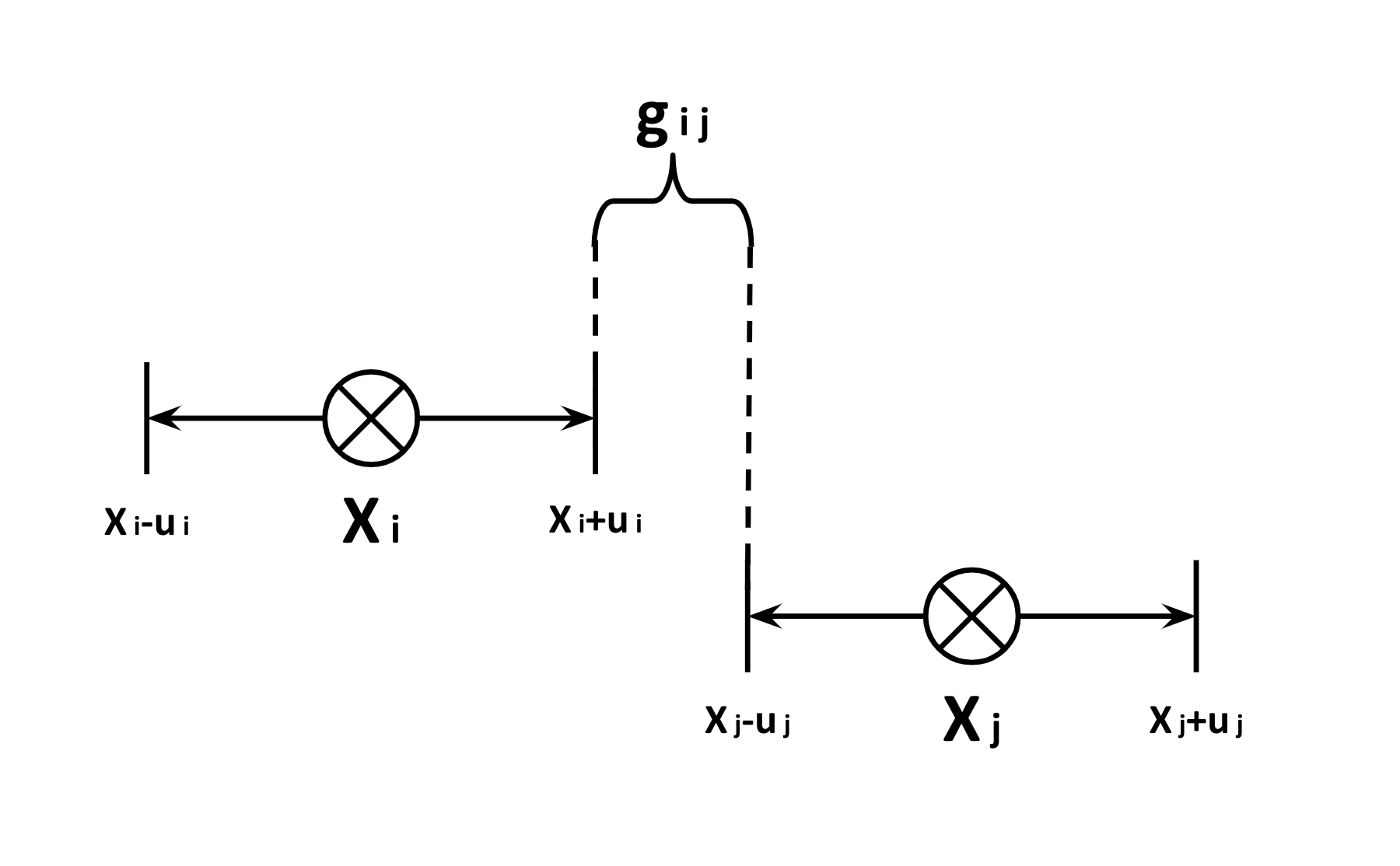}
    \caption{Illustration of non overlapping distance $g_{ij}$ for experts $i$ and $j$ with opinions $X_i$ and $X_j$ and uncertainties $u_i$ and $u_j$ respectively}
    \label{fig:relativedisagreement}
\end{figure}

Subtract the extent of the overlap $2u_i - g_{ij}$ to give the total disagreement:
\[g_{ij} - (2u_i - g_{ij}) = 2 (g_{ij} - u_i)\]
The RD between $i$ and $j$ is given by:
\[RD_{ij} = 2 (g_{ij} - u_i)/2u_i = (g_{ij}/u_i)-1\]
If $g_{ij} > u_i$, update the opinions and uncertainties with probability $\lambda$, where $\lambda$ is a parameter. 
\[x_j:=x_j+\mu RD_{ij}(x_i - x_j)\]
\[u_j:=u_j+\mu RD_{ij}(u_i - u_j)\]

\subsection{Markets and Traders}

The famous 18th-Century Scottish economist Adam Smith included a description of what he called \textit{The Invisible Hand} in his landmark book \cite{theInvisibleHand}; Smith used the term to embody the unintended positive effects of selfish behaviour in a market. This idea forms the basis for \textit{allocative efficiency}, sometimes thought as the ``fairness" of a market. Where utility is the measure of the usefulness a person gets from a product, the \textit{allocative efficiency} of a market is the total utility gained from trade, expressed as a percentage of the maximum possible utility to be gained.

Understanding the details of how selfish interactions among competitive traders in a market can give rise to desirable outcomes, such as efficient allocation of scarce resources between producers and consumers, has been a desire of economists ever since Adam Smith. A major step forward was taken by American economist Vernon Smith who in the late 1950s started a program of experimental studies of human traders interacting in markets under repeatable laboratory conditions -- a field that became known as {\em experimental economics}, the founding and growth of which resulted in Vernon Smith being awarded the Nobel Prize in Economics in 2002. Much of Smith's experimental work studied the dynamics of markets in which human traders, either {\em buyers} announcing bid-prices or {\em sellers} announcing ask-prices, interacted with one another via a market mechanism known as the {\em continuous double auction} (CDA) which is the basis of almost all of the world's major financial markets. In a CDA a buyer can announce a bid at any time and a seller can announce an offer at any time, and any buyer is free to accept an ask at any time while any seller is free to accept a bid at any time. 

In establishing experimental economics research, Vernon Smith had devised experimental CDA auctions for teaching purposes and later as a tool to observe how traders in a market act according to different specified conditions \cite{VernonSmith}. Vernon Smith and his fellow experimental economists focused entirely on the interactions among human traders in their market laboratories but in 1993, inspired by Vernon Smith's work, the economists Gode \& Sunder devised experiments to compare the allocative efficiency of minimally-simple automated trading systems against human traders. Gode \& Sunder's automated traders we so simple that they were, entirely justifiably, referred to as \textit{zero-intelligence} (ZI) traders. Most notably, in \cite{GodeandSunder} the authors describe the design of a ZI trader known as ZIC (for ZI-Constrained) which generated random bid or ask prices, subject to the single {\em budget constraint} that the prices generated should not lead to loss-making deals:
ZIC is constrained by a {\em limit price} and so draws its bid quote price from a uniform random distribution below the limit price, and its ask quote price from a uniform random distribution above the limit price.

To everyone's surprise the allocative efficiency scores of CDA markets populated by ZIC traders was demonstrated to be statistically indistinguishable from those of comparable CDA markets populated by human traders. Gode \& Sunder's result indicated to many people that the high intelligence of human traders was irrelevant within the context of a CDA-based market, and a research field formed, with various authors publishing details of automated trading systems that refined and extended the ZI approach. 

Often these early automated traders involved some means of making the trader adaptive, so that it could adjust its response to changing market conditions. As adaptivity to the environment is seen by some as a minimal signifier of intelligence, adaptive ZI-style automated trading agents became known as minimal-intelligence (MI) traders. 

Numerous variations on ZI/MI traders have been proposed to test the limits of their trading performance and to provide more human-like trader to test new trading strategies against. A notable work, which extended a MI trading strategy to enable the study of asset price bubbles and crashes, is \cite{Duffy}, discussed in more detail below.  

The primary contribution of this paper is to combine the Opinion Dynamics models with ZI/MI automated traders, creating a new class of automated trading strategies: ones that are still zero- or minimal- intelligence, but which also hold opinions. 

In the 27 years since Gode and Sunder published their seminal 1993 paper on ZIC, the field of agent-based computational economics (ACE) has grown and matured. For reviews of work in this field, see \cite{chen2018book,hommes_lebaron_2018}. ACE is a subset of research in agent-based modelling (ABM), which uses computational models of interacting agents to study various phenomena in the natural and social sciences: see \cite{ABM} for more details.

\subsection{The BSE Financial Exchange}

We used the \textit{BSE} open-source simulator of a contemporary financial exchange populated with a number of automated trading systems. The BSE project is open source and publicly available on Github, at:  \url{https://github.com/davecliff/BristolStockExchange} \cite{BSE}. 

BSE is a simulated CDA-based financial market, which is populated by a user-specifiable configuration of various automated-trader systems; it includes a number of predefined classes of automated trader each with unique trading strategies. 

BSE's implementation of a CDA, like real-world financial exchanges, requires buyers and sellers to submit bid and ask prices simultaneously and continuously onto an exchange mechanism that publishes the orders to a Limit Order Book, (\textit{LOB}), each order (each bid or ask) specifies a price and a quantity. A transaction will go through when a buyer's bid price and a seller's ask price are the same or 'cross', i.e. if a buyer's bid exceeds a seller's ask, or a seller's ask is less than a buyer's bid. When the transaction is complete, the orders have been filled hence they are removed from the \textit{LOB}. 
On a Limit Order Book (LOB), the bids and asks are stacked separately on ordered lists each sorted from best to worst: the best bid is the highest-priced one and the remaining bids are listed in decreasing-price order below it; the best ask is the lowest-priced one and the remaining asks are listed in ascending-price-order below it.


BSE comes with several types of ZI/MI automated traders built-in, including Gode \& Sunder's ZIC, and also Vytelingum's {\em AA} trader \cite{vytelingum2006} which was demonstrated by \cite{deluca_cliff_2011} to outpefrom human traders, so an experimental market can readily be set up and populated with some number of traders of each type. However BSE does not include the \textit{Near-Zero Intelligence} (NZI) trader-type introduced by \cite{Duffy}, so we created our own implementation of that and added it to BSE: the source-code for that implementation is available in our GitHub repository, the location of which was given in the footnote in Section 1. In the next section we describe NZI traders in more detail.   

\section{Near-Zero-Intelligence Traders}
\label{sec:NZItraders}

In \cite{Duffy}, NZI traders are defined to mimic the behaviour of traders in markets where asset prices bubble and crash, i.e.\ where the price of a tradeable asset rises quickly and falls precipitously. As the name implies, NZI traders are similar to Gode and Sunder's ZI traders but have some added features. The following is a summary of key aspects of NZI traders.

\subsection{The Weak Foresight Assumption}

Firstly, Duffy and Ünver define the \textit{weak foresight assumption} (WFA) which gives the traders knowledge that the trading session is coming to an end. This involves two variables: $\Bar{D}^T_t$ and $\pi_t$, both of which are explained further below.

A trading period is defined as 240 seconds where at the end of a trading period the traders earn a dividend per unit of the asset they own. The dividend amount is a random variable drawn from a uniform distribution with support: ${d_1, d_2, d_3, d_4}$ where $\{0\leq d_1<d_2<d_3<d_4\}$. Hence the expected dividend is given by:
\[\Bar{d} = \frac{1}{4}\sum_{i=1}^{4} d_i\]

At the start of each simulation of $T$ trading periods, a trader $i$ has a balance of $x_i$ and owns a number $y_i$ of units of the tradeable asset. Before the first trading period, $t=1$, we have the equation:

\[x_i + \Bar{D}^T_1 y_i = c\]
where $c$ is a constant for all $i$. 

During the simulation of the market sessions, $\Bar{D}^T_t$ decreases as $t\rightarrow T$. It represents the fundamental market price or the default value of the asset at period $t$ which earns zero profit. It is calculated by the equation: 
\[\Bar{D}^T_t = \Bar{d}(T-t+1) + \Bar{D}^T_{T+1}\]
$\Bar{D}^T_t $ is a value that decreases by $\Bar{d}$ each trading period $t$, this makes up the first part of the WFA.

The second part of the WFA is $\pi_t$, the probability of a trader being a buyer in trading period $t$. It is given by the equation:
\[\pi_t = max\{0.5-\varphi t, 0\}\]
where
$\varphi \in [0,{0.5}/{T})$.
Since $0 \leq \varphi < \frac{0.5}{T}$ then $0 < \pi_t \leq 0.5$, and as $t \rightarrow T$, the probability of a trader being a buyer decreases over time; therefore traders are less likely to buy as time goes by. The combination of a reduction in tendency to buy, caused by $\pi_t$, and a decrease in the default value of the asset, $\Bar{D}^T_t$, results in traders having a ``weak" awareness of the future hence, the name ``weak foresight assumption".

\subsection{The Loose Budget Constraint}

In \cite{GodeandSunder}, their ZIC trader has a \textit{no loss constraint}. That constraint on ZIC traders forces them to buy and sell at prices bounded by the intrinsic value, and transacting at that price would not result in asset price inflation.

In contrast to Gode and Sunder's work, \cite{Duffy} propose a \textit{``loose" budget constraint}: if trader $i$ is a seller and has an asset, submit an ask price; and if trader $i$ is a buyer and has sufficient cash balance, submit a bid price:

\begin{algorithmic}
\IF{trader $i$ is a seller \AND trader $i$ has an asset} 
\STATE {submit ask} \ELSIF{trader $i$ is a buyer} \STATE{submit min(balance, bid)} \ENDIF
\end{algorithmic}

\subsection{The ``Anchoring Effect"}

Another departure from \cite{GodeandSunder} is that Duffy \& Ünver's NZI traders are not entirely \textit{zero-intelligence}. In fact they have knowledge of the mean transaction price from the previous trading period, denoted $\Bar{p}_{t-1}$, which is used to calculate the trader's initial quote price in a trading period -- thus the trader's quote price is to some extent ``anchored'' by the previous period's prices. In the first session, $\Bar{p}_{t-1}=0$, and the traders submit low quote prices.

\subsection{Formal Specification}

Simulations involve $T$ market periods or sessions, $t \in [1,T]$, and within each iteration of each market session a trader $i$ is chosen to submit an order in sequence $S$, $s \in S$. The uniform random variable $u^i_{t,s}$ is calculated using $\Bar{D}^T_t$ via:
\[u^i_{t,s} \in [\underline\epsilon_t, \Bar\epsilon_t]\]
where 
$\underline\epsilon_t = 0$,
$\Bar\epsilon_t = k\Bar{D}^T_t$
and $k > 0$ is a parameter. The upper bound of $u^i_{t,s}$, $\Bar\epsilon_t$, will decrease over time since $\Bar{D}^T_t$ decreases. Therefore the range for $u^i_{t,s}$ becomes smaller and with an average of $\frac{1}{2}k\Bar{D}^T_t$, the value of $u^i_{t,s}$ should decrease.

If a trader is a seller then offer the ask price $a^i_{t,s}$,
\[a^i_{t,s} = (1-\alpha) u^i_{t,s} + \alpha \Bar{P}_{t-1},\]
where $\alpha \in (0,1)$ is a constant parameter. Using the \textit{loose budget constraint} so a buyer can only offer as much money as they possess, if a trader is a buyer then offer the bid price $b^i_{t,s}$,
\[b^i_{t,s} = min\{(1-\alpha) u^i_{t,s} + \alpha \Bar{P}_{t-1}, x^i_{t,s}\}\]


\begin{figure}[!h]
    \centering
    \includegraphics[width=0.8\linewidth]{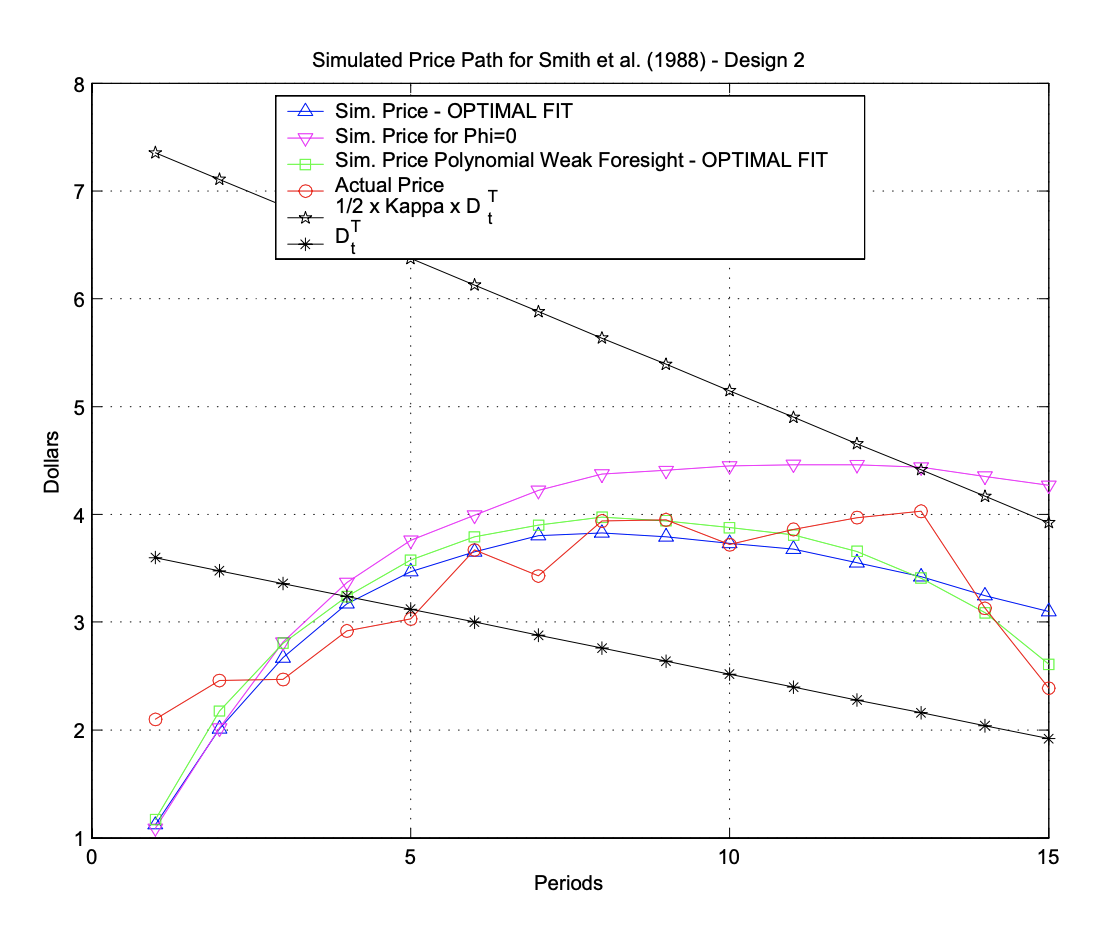}
    \caption{Comparison of mean transaction price path in the simulations and actual data from \cite{Duffy}}
    \label{fig:duffy}
\end{figure}

The combination of a decreasing $\Bar{D}^T_t$ value and an anchoring to the mean transaction price of the previous trading period $\Bar{P}_{t-1}$ results in a humped shape pattern in the transaction history. This hump is the model's endogenous rise in price, i.e.\ the `bubble', followed by a fall or `crash'. The mean transaction price per trading period increases initially due to the high $\Bar{D}^T_t$ value which increases the bid and ask prices above the previous mean transaction price $\Bar{P}_{t-1}$. Eventually as the value of $\Bar{D}^T_t$ decreases, the mean transaction price levels out closer to $\alpha \Bar{P}_{t-1}$ which is less than or equal to $\Bar{P}_{t-1}$.

\section{Opinionated Traders}
\label{sec:opinionatedtraders}

We introduce a new variation on the ZIC trader model, from \cite{GodeandSunder}, called the Opinionated-ZIC (i.e., OZIC) trader, that submits quote-prices affected by its opinion.

\begin{figure}[!h]
     \centering
     \begin{subfigure}[b]{0.6\linewidth}
         \centering
         \includegraphics[width=0.6\textwidth]{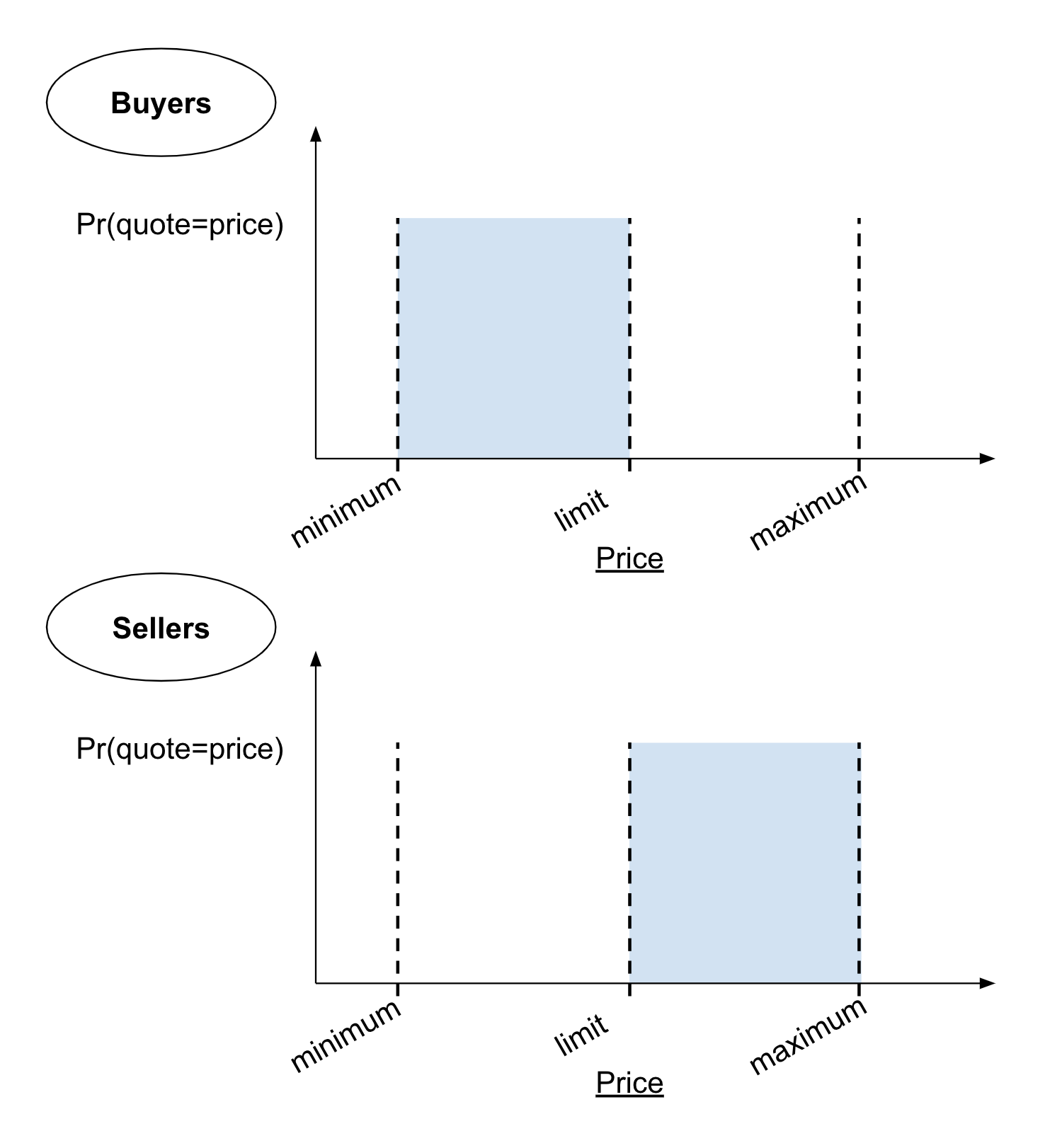}
         \caption{Quote price range of ZIC traders\vspace*{2em}}
         \label{fig:OZIC1}
     \end{subfigure}
    \vspace*{1em}
     \begin{subfigure}[b]{0.6\linewidth}
         \centering
         \includegraphics[width=0.6\textwidth]{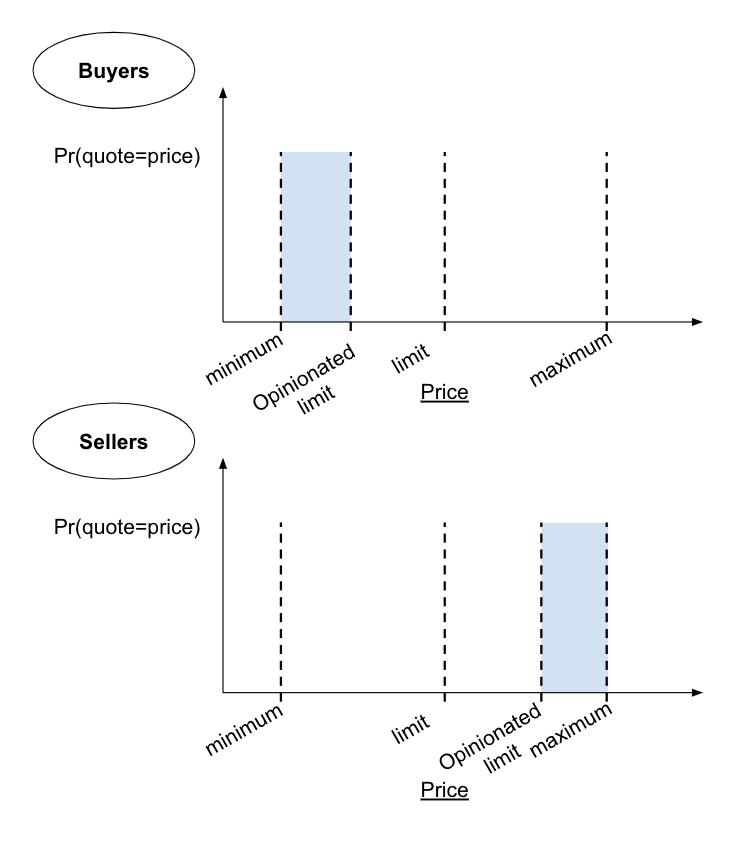}
         \caption{Quote price range of OZIC traders}
         \label{fig:OZIC2}
     \end{subfigure}
     \vspace*{0.5em}
    \caption{Diagrams of quote price range for Gode \& Sunder's Zero Intelligence Constrained (ZIC) Traders in \ref{fig:OZIC1} and for our Opinionated-ZIC (OZIC) Traders in \ref{fig:OZIC2}. The shaded region represents the uniform distribution that the traders' quote prices are  drawn from.}
    \label{fig:OZIC}
\end{figure}
The BSE simulator \cite{BSE} contains an implementation of the ZIC trader, which has knowledge of the Limit Order Book (LOB), it sets its minimum quote price to the worst bid on the LOB, its maximum quote price to the best ask price on the LOB, and its limit price to that specified by the customer order currently being worked on. 
If the ZIC trader is a buyer then it submits orders with a quote price generated from a random draw between the minimum quote price and the limit price. Otherwise, if the ZIC trader is a seller then it submits orders with a quote price generated from a random draw between the limit price and the maximum quote price. The quote price distribution for ZIC traders are illustrated in Figure \ref{fig:OZIC1}, with the buyers' quote price distribution on the left and the sellers' quote price distribution on the right.

The \textit{Opinionated Zero-Intelligence-Constrained} (OZIC) trader model submits quote prices that vary according to its opinion. If the OZIC trader is a buyer and its opinion is negative then it submits a low bid, and if its opinion is positive then it submits a bid that is higher but still capped at its limit price. On the other hand if the OZIC trader is a seller and its opinion is negative then it submits a low ask, and if its opinion is positive then it submits a high ask. This models the idea that traders will submit quote prices close to what they believe the actual value of the stock to be, and if a traders holds a positive opinion of the stock they would believe the value of the stock to be greater than a trader holding a negative opinion of the stock.

As illustrated in Figure \ref{fig:OZIC2}, the quote price range for OZIC buyers are between the minimum price and their \textit{opinionated limit}, and the quote price range for OZIC sellers are between their opinionated limit and the maximum price.

If the OZIC trader $i$ is a buyer then calculate the opinionated limit $OL_i$ by:
\[OL_i  = f(x) = \frac{L (1 + x_i) + \underline{M} ( 1 - x_i )}{2},\]
where $L$ is the limit price, $\underline{M}$ is the minimum price, and $x_i$ is the opinion of OZIC trader $i$: this gives $f(-1)=\underline{M}$; $f(0)=\frac{L + \underline{M}}{2}$: and $f(1)=L$. Then generate a bid quote price as a random draw from the interval $[\underline{M}, OL_i]$.

If the OZIC trader $i$ is a seller then calculate the opinionated limit $OL_i$ by:
\[OL_i  = f'(x) = \frac{L (1 - x_i) + \Bar{M} ( 1 + x_i )}{2},\]
where $L$ is the limit price, $\Bar{M}$ is the maximum price, and $x_i$ is the opinion of OZIC trader $i$: this gives $f'(-1)=L$; $f'(0)=\frac{L + \Bar{M}}{2}$; and $f(1)=\Bar(M)$. Then bid quote prices are generated as a random draw from the interval $[OL_i, \Bar{M}]$.

\subsection{Opinionated NZI Traders}

We also introduce here an {\em Opinionated Near-Zero-Intelligence} (ONZI) trader based on the \textit{near-zero-intelligence} (NZI) trader model of  \cite{Duffy}. The ONZI trader model offers the possibility of price bubbles dependent on the prevailing opinions of the population, i.e.\ if the opinions are mostly positive then the bubble should be greater than if the opinions were mostly negative.  

\subsection{Recreating NZI trader model}

Duffy \& Utku Ünver's NZI trader model uses a random component $u_{t,s}^i$, given by 
$u_{t,s}^i \in [0, k\Bar{D}^T_t],$
where $i$ is the index of the trader, $t$ is the current trading period out of $T$ periods, $s$ is the order of the trader in the sequence that the traders submit orders, $k$ is a constant parameter, and $\Bar{D}^T_t$ is the default value of the asset. The ask price $a_{t,s}^i$ is calculated using $u_{t,s}^i$ as described in Section~\ref{sec:NZItraders}.

In \cite{Duffy}, optimal parameter values were calibrated to best match their simulated data with the data collected from experiments with human traders. The values are as follows: $k^*=4.0846,$ $\alpha^*=0.8480,$ $\phi^*=0.01674,$ and $S^*=5$. We use the optimised parameter values $k^*$ and $\alpha^*$ hereafter, however we have not used $\phi^*$ because in our work the buyers and sellers do not change specification and we have not used $S^*$ as small values of $S$ do not show opinion convergences in large populations very well. The ask and bid price of traders are calculated in such a way that they require the default value $\Bar{D}^T_t$ of the asset and the mean transaction price of the previous trading period $\Bar{P}_{t-1}$. 
To get the default value of $\Bar{D}^T_t$ for each trading period $t$, the expected dividend amount $\Bar{d}$ is calculated by the average of dividends $[0,1,2,3]$ which is $1.5$ and the final value is set $\Bar{D}^T_{T+1}=40$. These values form a similar gradient for $\Bar{D}^T_t$ over time to that shown in \cite{Duffy}. 

\subsection{Opinionated Limit}

\begin{figure}[!h]
     \centering
     \begin{subfigure}[b]{\linewidth}
         \centering
         \includegraphics[width=0.4\textwidth]{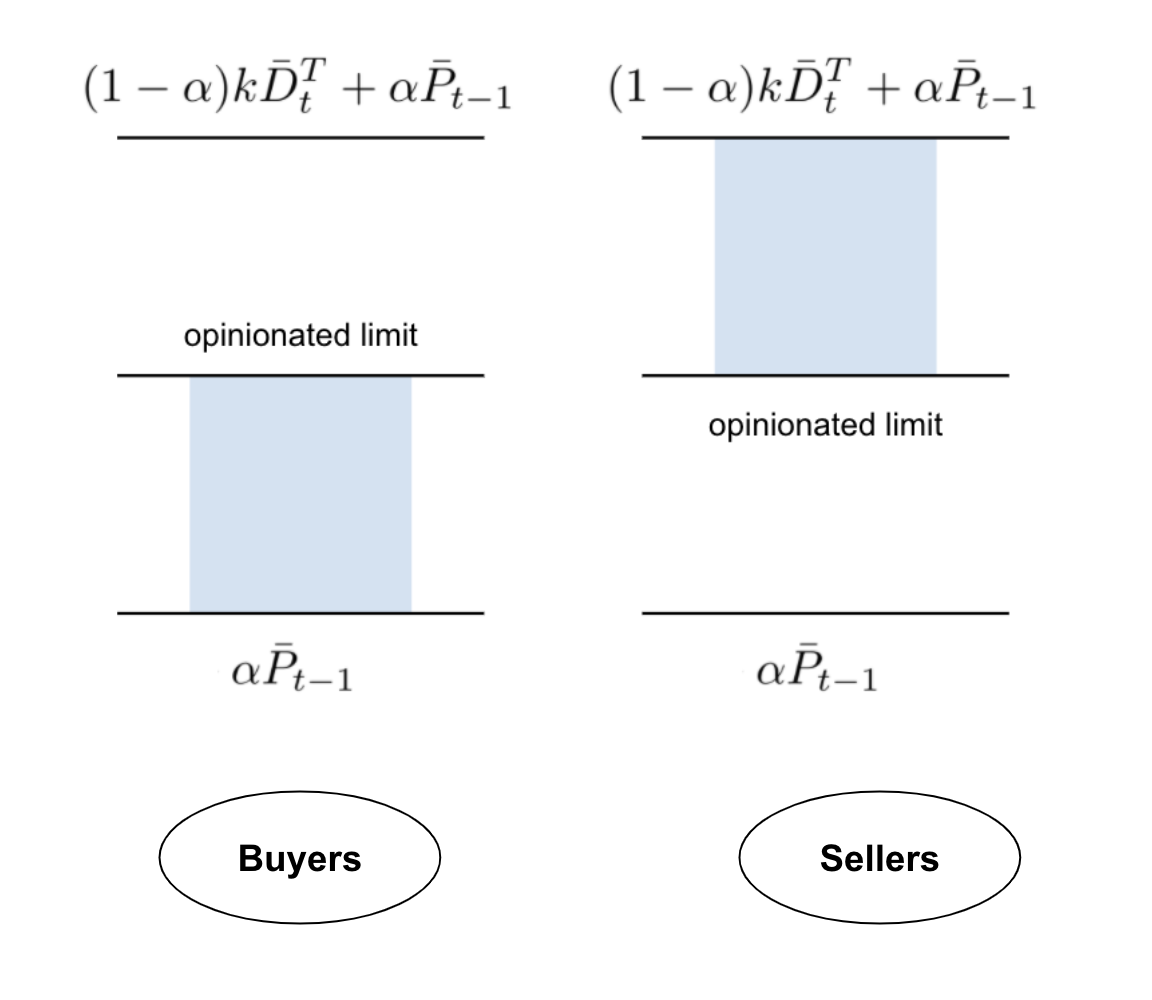}
         \caption{\vspace*{2em}}
         \label{fig:ONZIC1}
     \end{subfigure}
     \hfill
     \begin{subfigure}[b]{\linewidth}
         \centering
         \includegraphics[width=0.4\textwidth]{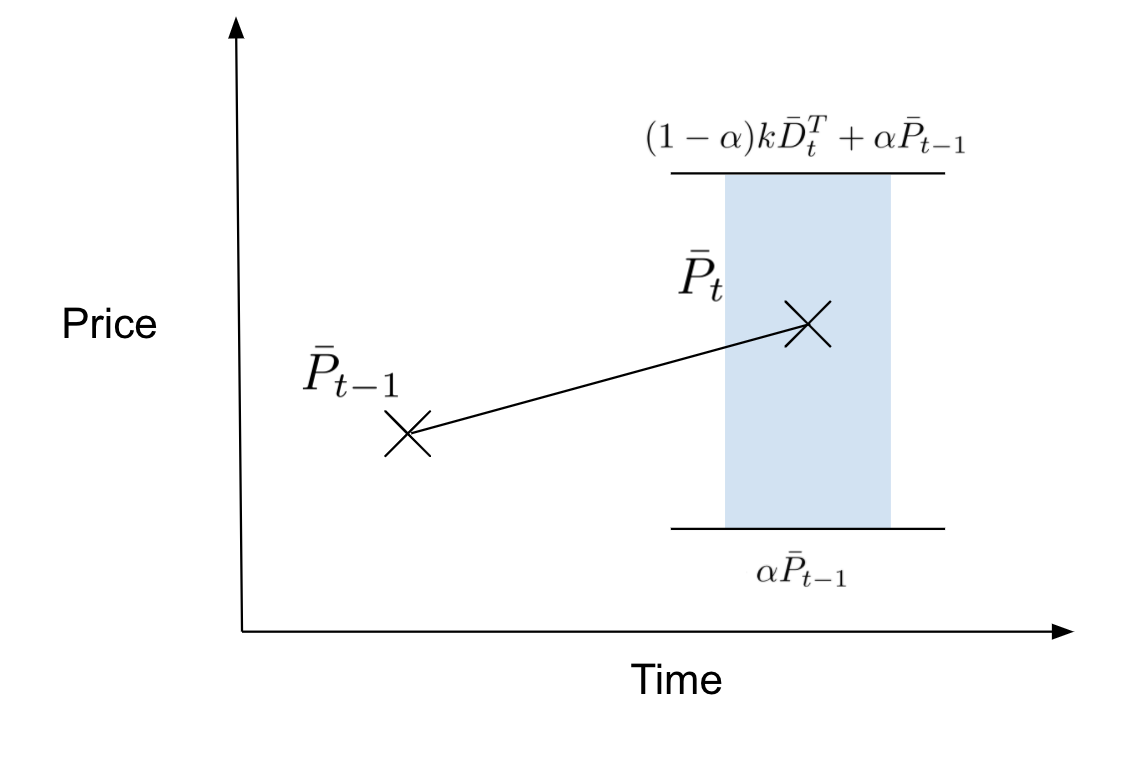}
         \caption{}
         \label{fig:ONZIC2}
     \end{subfigure}
    \caption{Diagram of quote price range for Opinionated near-zero-intelligence (ONZI) Traders in \ref{fig:ONZIC1} and an illustration of the possible range for the mean transaction price $\Bar{P}_t$ of trading period $t$ in relation to the previous mean transaction price $\Bar{P}_{t-1}$ in \ref{fig:ONZIC2}.}
    \label{fig:ONZIC}
\end{figure}

We created an \textit{opinionated limit} to integrate trader opinions with the NZI strategies. Similarly to the opinionated limit calculation in our OZIC trader model, the opinionated limit of the ONZI trader model can be calculated from between $\alpha \Bar{P}_{t-1}$ and $(1-\alpha)k\Bar{D}^T_t + \alpha \Bar{P}_{t-1}$, as shown in Figure \ref{fig:ONZIC1}, because the maximum $u_{t,s}^i$ value is $k\Bar{D}^T_t$.
So for an ONZI trader $i$, with opinion $x_i$, the opinionated limit $OL_i$ is calculated by:
\[OL_i = \frac{(1-\alpha)(k\Bar{D}^T_t + \alpha \Bar{P}_{t-1})(1+x_i)+ (\alpha \Bar{P}_{t-1})(1-x_i)}{2}\]
This form is closest to that of OZIC traders but is easier to read when expressed in terms of the \textit{opinionated uncertainty} $OU_{t,s}^i$, based on the definition of $u_{t,s}^i$, which is
given by:
\[OU_{t,s}^i\in [0, \frac{1}{2} k \Bar{D}^T_t (1+x_i)]\]

Then the quote price $a^i_{t,s}$ is calculated by:
\[a^i_{t,s} = (1-\alpha)OU^i_{t,s} + \alpha \Bar{P}_{t-1}\]


The effect of the opinionated uncertainty $u^i_{t,s}$ is illustrated in Figure \ref{fig:ONZIC2}, where the value of $\Bar{P}_t$ is the mean transaction price for trading period $t$. During trading period $t$, every trader will submit quotes between $\alpha \Bar{P}_{t-1}$ and $(1-\alpha)k\Bar{D}^T_t + \alpha \Bar{P}_{t-1}$ so if there are $n$ transactions that take place at the maximum $(1-\alpha)k\Bar{D}^T_t + \alpha \Bar{P}_{t-1}$ then the average $\Bar{P}_{t}$ will be:
\[\frac{1}{n}\sum^{n}((1-\alpha)k\Bar{D}^T_t + \alpha \Bar{P}_{t-1}) = (1-\alpha)k\Bar{D}^T_t + \alpha \Bar{P}_{t-1}\]
Similarly if all transactions in trading period $t$ occur at the minimum $\alpha \Bar{P}_{t-1}$, then the average $\Bar{P}_{t}$ will be:

\[\frac{1}{n}\sum^{n}(\alpha \Bar{P}_{t-1}) = \alpha \Bar{P}_{t-1}\]

The shaded region in Figure \ref{fig:ONZIC2} represents the range that $\Bar{P}_t$ can be in, i.e. between $\alpha \Bar{P}_{t-1}$ and $(1-\alpha)k\Bar{D}^T_t + \alpha \Bar{P}_{t-1}$. 
The value of $\Bar{D}^T_t$ will decrease hence the range for $\Bar{P}_{t}$ decreases however will roughly remain centered.
In contrast, a population of ONZI traders will submit high quote prices,  close to the maximum, when they hold positive opinions and will submit low quote prices, close to the minimum, when they hold negative opinions.

\section{Results}
\label{sec:results}
\subsection{OZIC Traders}
\subsubsection{Baseline Results}

\begin{figure}[!h]
  \centering
  \includegraphics[trim=145 20 20 10, clip,width=\linewidth]{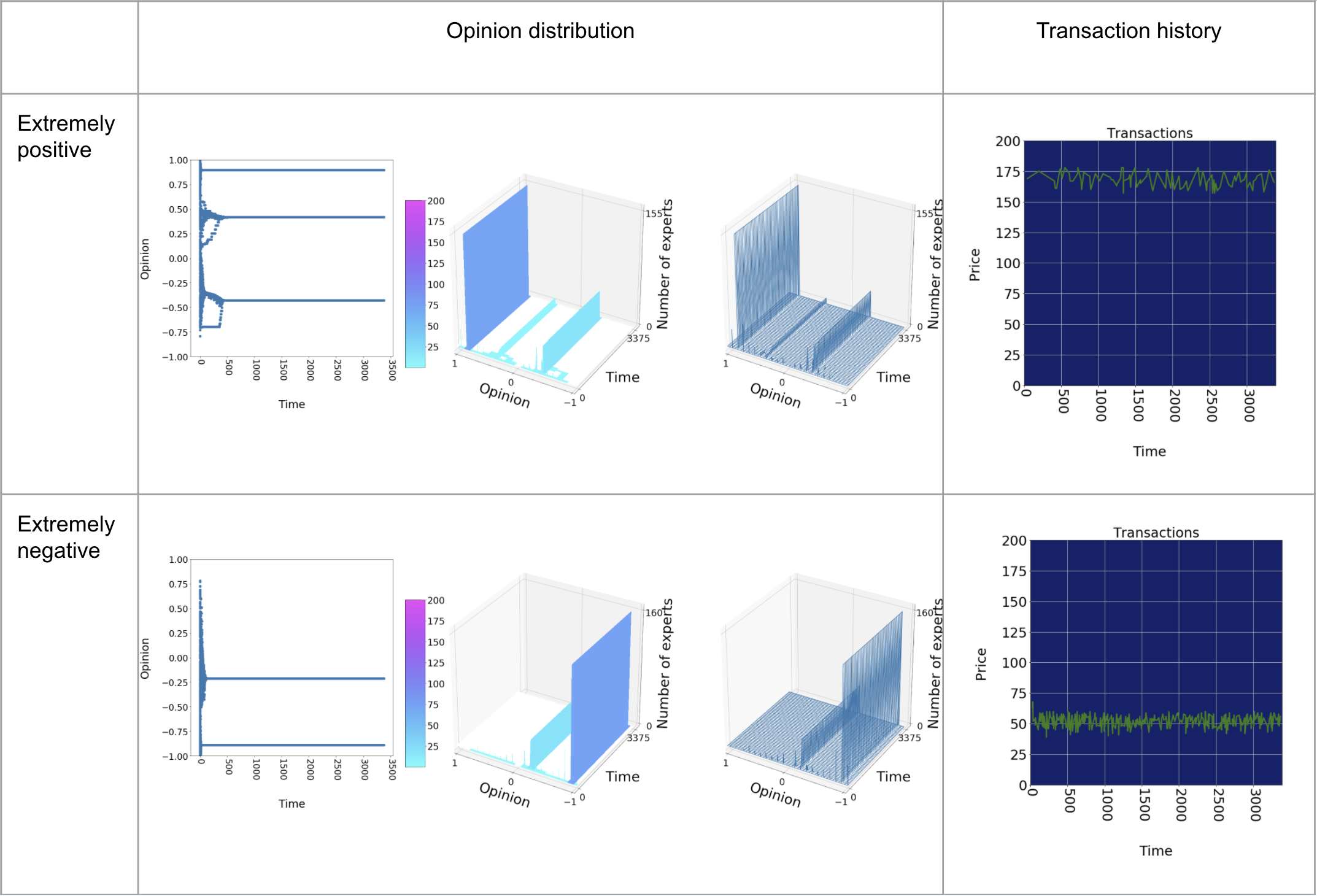}
  \caption{OZIC traders with extreme opinions. Upper row of plots is for traders with extremely positive opinions; lower row is for traders with extremeley negative opinons. The plot at far left shows the convergence of opinion values in the population over time, in the 2D style used by \cite{deffuant2002} among others -- the population converges to a situation where all traders hold one of three opinions; the two central plots display the same opinion-distribution data as 3D plots (heatmap-colored on the left; uncoloured on the right), which gives a better indication of the number of traders that hold each converged-upon opinion. The dark-background plot at far right in each row os the transaction-price time series from this experiment.}
  \label{fig:extremesOZIC}
\end{figure}

The more useful results are in the extreme cases of opinion distribution, i.e. when all the traders hold extremely positive opinions or negative opinions. In Figure \ref{fig:extremesOZIC}, we have shown the effects of extremely positive opinion distribution on the transaction history which is quite high, whereas for an extremely negative opinion distribution the transaction history shows very low prices.
The results use the RA model with $pe=0.5$ and $w=0.5$, and a function that specifies the distribution of extremists.

\begin{figure}[!h]
    \centering
    \includegraphics[width=0.4\linewidth]{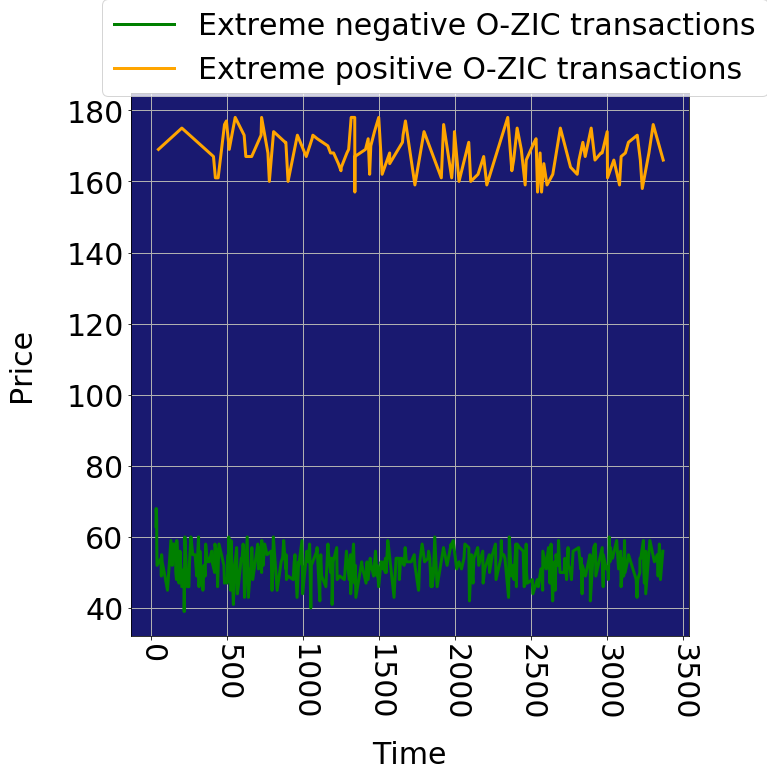}
    \caption{Comparison of OZIC trader transaction histories with extremely negative and positive opinions}
    \label{fig:compOZIC}
\end{figure}

In Figure \ref{fig:compOZIC}, we have plotted the transaction histories of OZIC traders with extremely positive opinions, in orange, and extremely negative opinions, in green. When compared this way it is clear that the traders with extremely positive opinions trade at much higher prices than traders with extremely negative opinions.  

\subsubsection{Extreme Opinion Shift}

\begin{figure}[!h]
    \centering
    \includegraphics[trim=145 20 20 10, clip,width=0.6\linewidth]{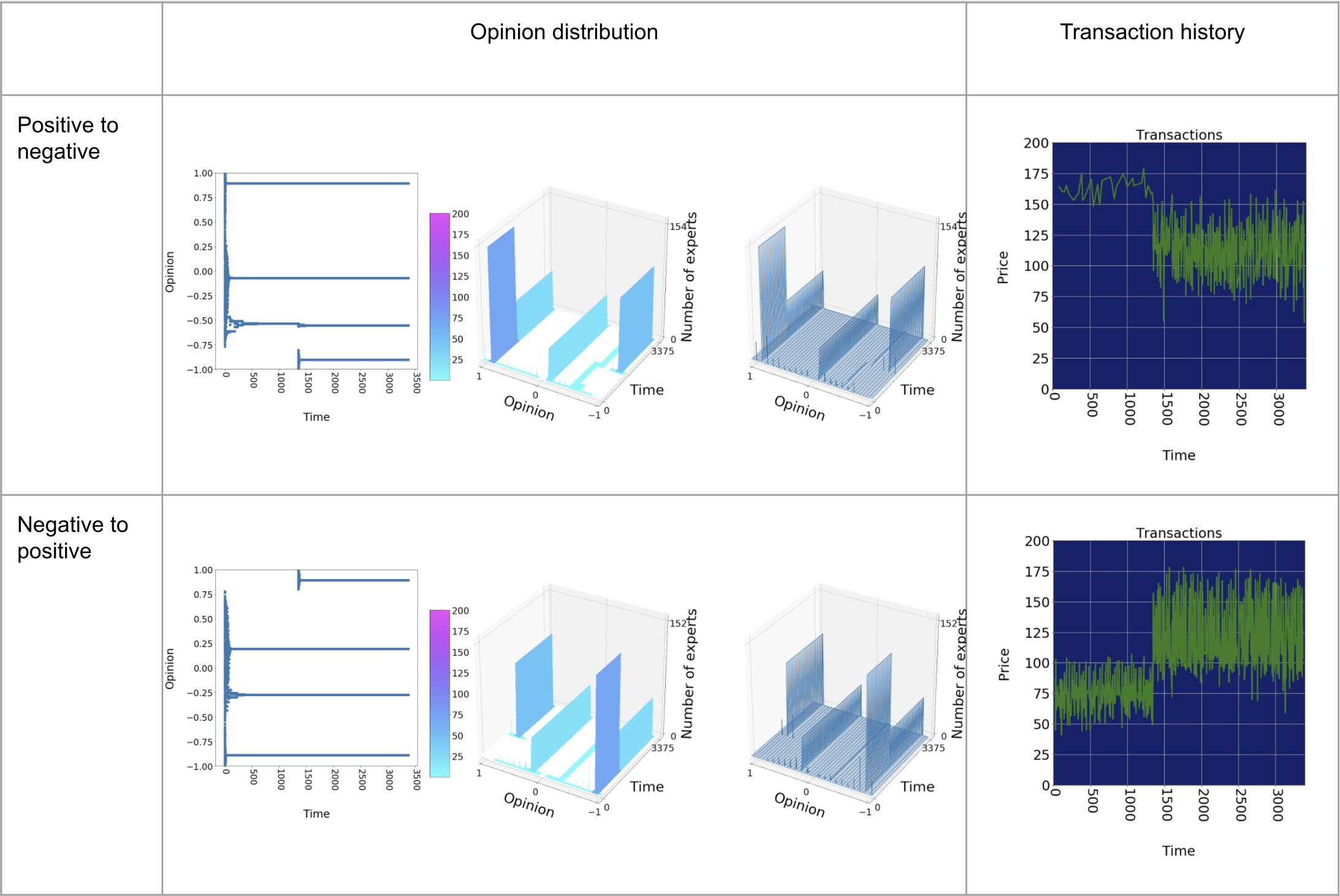}
    \caption{OZIC traders with extreme shifts in opinion at the start of Period 6; format as for Figure~\ref{fig:extremesOZIC}.}
    \label{fig:OZICshifts}
\end{figure}

We initialise a given proportion of extremists to be extremely positive or negative initially and switch them to the polar opposite opinion half way through the duration of the simulation. Figure \ref{fig:OZICshifts} shows the results for a population of 100 OZIC buyers and 100 OZIC sellers using the RA model with proportion of extremists $pe=0.5$, confidence factor $\mu=0.5$, and uncertainty in the range $[0.2, 2.0]$. 

The results show a clear change in mean transaction price in relation to opinion distribution. For a positive to negative opinion shift, the traders start selling and buying at high prices and after $t=1350$ drastically shift to lower prices. Similarly for a negative to positive opinion shift, the traders begin trading at low prices and after $t=1350$ trade at higher prices. 

\subsection{ONZI Trader Results}

\subsubsection{Baseline Results}

The same rationality for testing the extreme opinion distributions for ONZI traders applies to testing ONZI traders. With extremely positive opinions, the shape of the transaction history peaks higher and has a greater initial gradient than that of ONZI traders with extremely negative opinions. ONZI traders with extremely negative opinions show a shorter hump shaped pattern than the ONZI traders with extremely positive opinions. 

\begin{figure}[!h]
    \centering
    \includegraphics[trim=145 20 20 10, clip,width=0.6\linewidth]{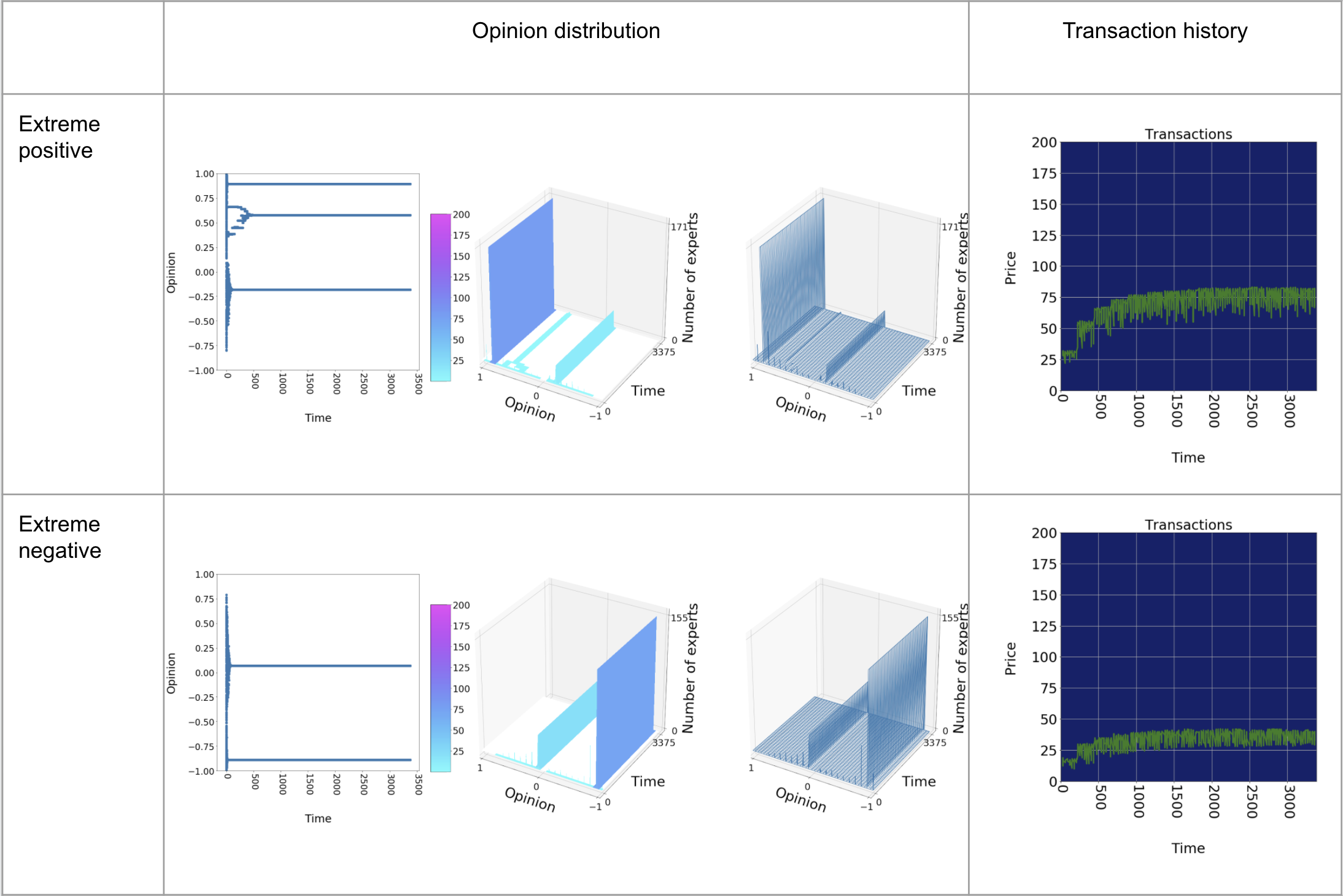}
    \caption{ONZI trader transaction histories with extreme positive and negative opinions; format as for Figure~\ref{fig:extremesOZIC}.}
    \label{fig:ONZICextreme}
\end{figure}

\begin{figure}[!h]
    \centering
    \includegraphics[trim=250 10 10 70, clip,width=8cm]{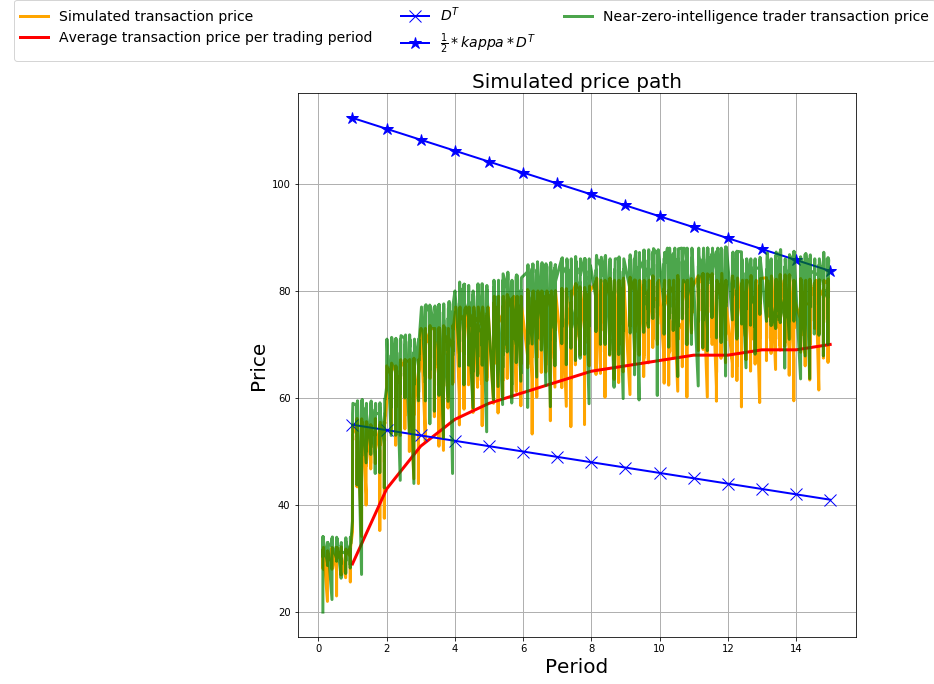}
    \caption{ONZI trader transaction history with extremely positive opinions; compared to the original NZI results shown in Figure \ref{fig:duffy}. Yellow lines show transaction history of traders with extreme positive opinions; green lines are baseline comparison; red line shows mean transaction price. }
    \label{fig:ONZICpospath}
\end{figure}

\begin{figure}[!h]
    \centering
    \includegraphics[trim=250 10 10 70, clip, width=8cm]{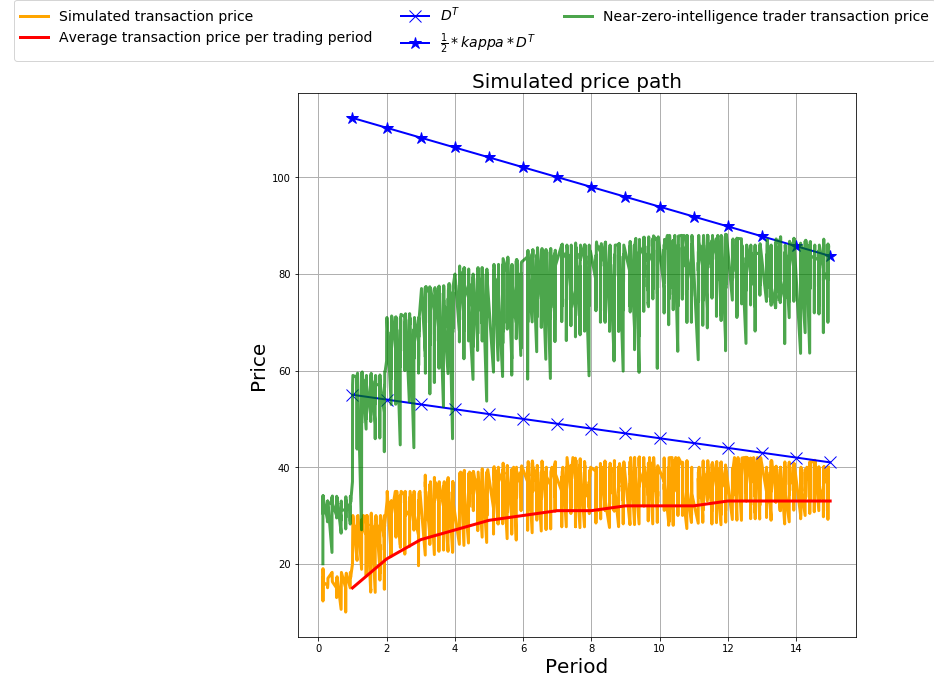}
    \caption{ONZI trader transaction history with extremely negative opinions; compared to the original NZI results as shown in Figure \ref{fig:duffy}. Color-coding of lines is as for Figure \ref{fig:ONZICpospath}.}
    \label{fig:ONZICnegpath}
\end{figure}

In Figures \ref{fig:ONZICpospath} and \ref{fig:ONZICnegpath}, inspired by a graph in \cite{Duffy}, we have plotted the transaction histories of the ONZI trader, in orange, against an ordinary \textit{near-zero-intelligence} (NZI) trader's results, in green. We have also plotted $\Bar{D}^T$ over time and $1/2 \kappa \Bar{D}^T$  over time to illustrate the effect it has on the transaction price over time. The average transaction price per trading period is also shown to encapsulate the overall behaviour of the market trends, in red. The simulated data for NZI traders, in green, tapers off and does not crash because we are not using a decreasing proportion of buyers in the population.

The transaction price data for ONZI traders with extremely positive opinions is very close to the simulated transaction history of \textit{near-zero-intelligence} traders, as shown in Figure \ref{fig:ONZICpospath}. On the other hand, the transaction price data for ONZI traders with extremely negative opinions is much lower than the simulated transaction history of \textit{near-zero-intelligence} traders, as shown in Figure \ref{fig:ONZICnegpath}.

\subsubsection{Extreme Opinion Shift}

\begin{figure}[!h]
    \centering
    \includegraphics[trim=145 20 20 10, clip,width=0.6\linewidth]{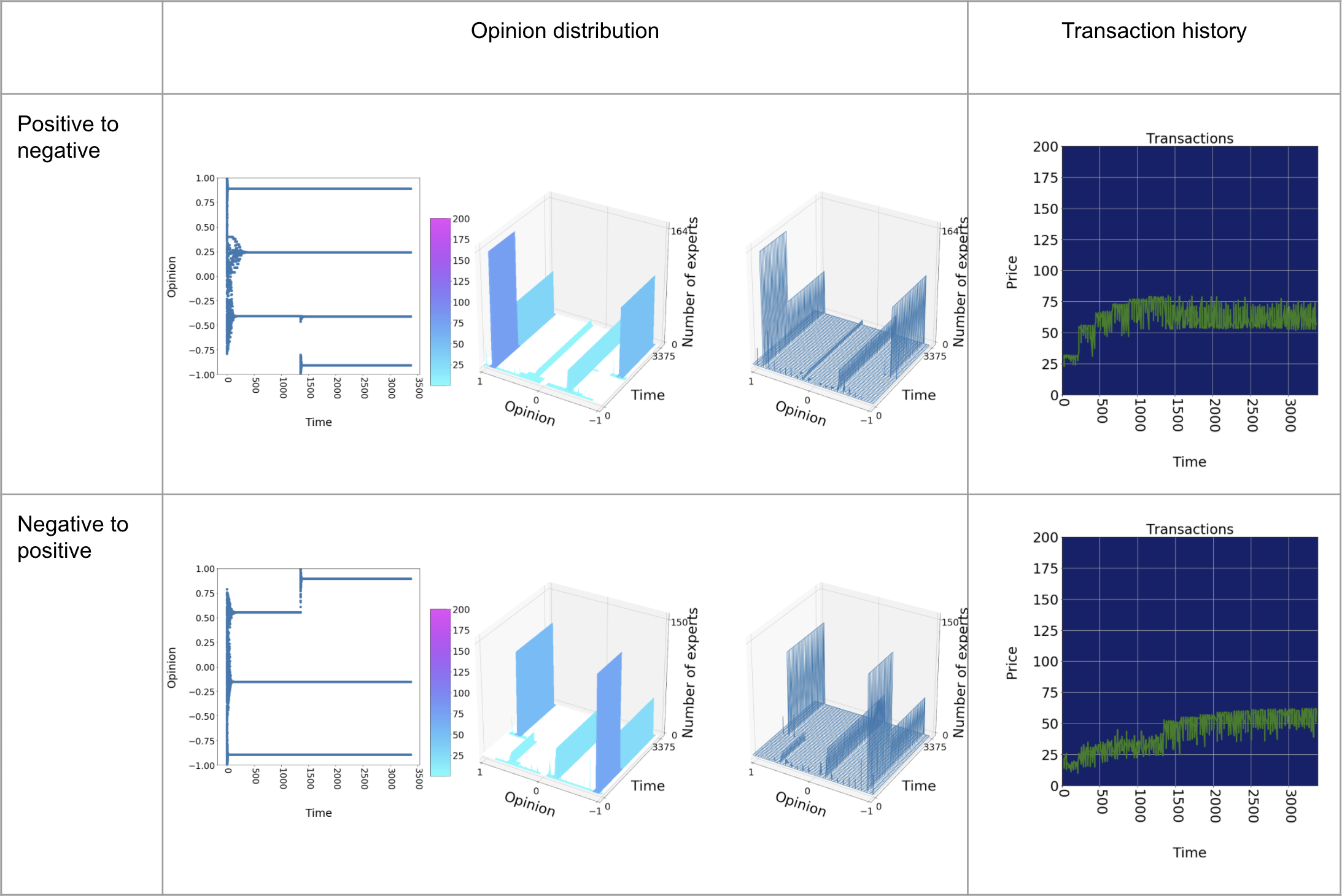}
    \caption{ONZI extreme opinion shifts; format as for Figure~\ref{fig:extremesOZIC}.}
    \label{fig:ONZICshifts}
\end{figure}

\begin{figure}[!h]
    \centering
    \includegraphics[trim=250 10 10 70, clip,width=0.6\linewidth]{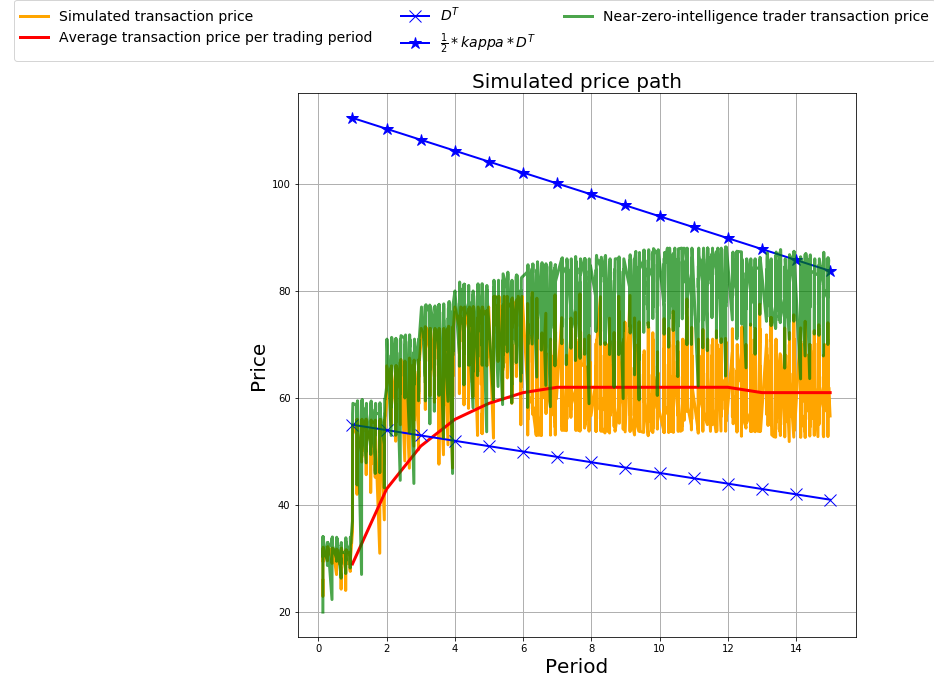}
    \caption{ONZI traders with extremely positive opinions drastically shifting to negative opinions at the start of Period 6. Color-coding of lines is as for Figure~\ref{fig:ONZICpospath}.}
    \label{fig:ONZICposneg}
\end{figure}

\begin{figure}[!h]
    \centering
    \includegraphics[trim=250 10 10 70, clip,width=0.6\linewidth]{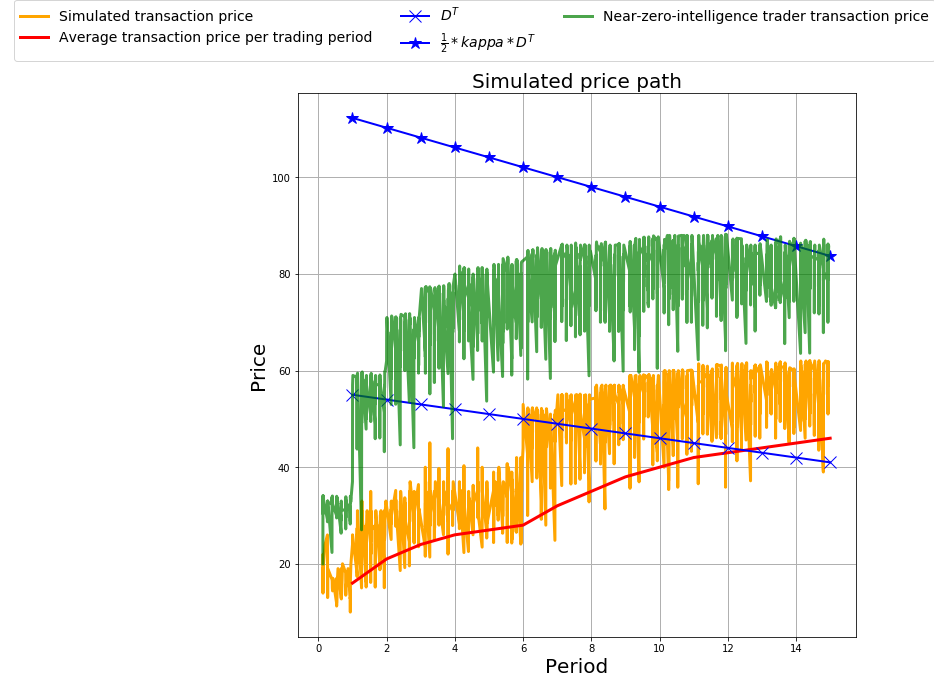}
    \caption{ONZI traders with extremely negative opinions drastically shifting to positive opinions at the start of Period 6.  
    Color-coding of lines is as for Figure \ref{fig:ONZICpospath}.
    }
    \label{fig:ONZICnegpos}
\end{figure}

Figure \ref{fig:ONZICshifts} shows ONZI traders with extremely positive opinions until half way through the simulation, i.e. $t=1350$, when the opinions shift to extremely negative, and vice versa. 
The opinion dynamics model used is RA with confidence factor $\mu=0.5$ and proportion of extremists $pe=0.5$ for both initializations of extremists.
Similarly to the results in Figures \ref{fig:ONZICposneg} and \ref{fig:ONZICnegpos}, we have plotted the transaction histories of ONZI traders with drastically shifting opinion distributions against the ordinary NZI traders, the default value $\Bar{D}^T$, the expected uncertainty $1/2 \kappa \Bar{D}^T$, and the mean transaction price per trading period. The mean transaction price per trading period, in red, is a useful indicator of the trends generated from the opinion distribution, as the average transaction price over time increases and decreases according to positive and negative opinions respectively.

\section{\uppercase{Conclusions}}
\label{sec:conclusion}

\noindent In this paper we have described what we believe to be the first ever system that integrates ideas from opinion dynamics into well-established trader-agent models, and in doing so we have created the first platform for the experimental exploration of agent-based models of narrative economics. In his seminal work on narrative economics, Nobel-Laureate Robert Shiller argues for a program of empirical research, gathering data on the stories, the narratives, that humans tell each other about economic affairs, which shape and change their opinions about future economic events, and where those opinions are themselves also significant factors in the dynamics of economic affairs. Our work opens up an experimental approach that is complementary to the one proposed by Shiller: using our platform, experimentalists can now also run agent-based simulations to better understand the dynamic interplay between opinions, expressions of those opinions, and subsequent economic outcomes.   

\clearpage 

\section*{\uppercase{Acknowledgements}}

\noindent The work described here was orally presented in October 2020 at an international conference on Zero- and Minimal-Intelligence Trading Agents held virtually at the Yale School of Management, Connecticut, USA. We are grateful to the participants of that meeting for their insightful questions and comments, and for awarding this work the Best Student Paper prize.

\bibliographystyle{apalike}
{\small
\bibliography{LomasCliff}}

\begin{thebibliography}{}

\bibitem[Chatterjee and Seneta, 1977]{chatterjee}
Chatterjee, S. and Seneta, E. (1977).
\newblock {Towards Consensus: Some Convergence Theorems on Repeated Averaging.}
\newblock {\em Journal of Applied Probability}, 14(1):88--97.

\bibitem[Chen, 2018]{chen2018book}
Chen, S.~H. (2018).
\newblock {\em Agent-based computational economics: How the idea originated and
  where it is going}.
\newblock Routeledge.

\bibitem[Cliff, 2018]{BSE}
Cliff, D. (2018).
\newblock {BSE : A Minimal Simulation of a Limit-Order-Book Stock Exchange.}
\newblock {\em {Proc.\ European Modelling and Simulation Symposium}}, pages
  194--203.

\bibitem[Cooks and Heppenstall, 2011]{ABM}
Cooks, A. and Heppenstall, A. (2011).
\newblock {Introduction to Agent-Based Modelling}.
\newblock {\em {Agent-Based Models of Geographical Systems}}, pages 85--105.

\bibitem[De~Luca and Cliff, 2011]{deluca_cliff_2011}
De~Luca, M. and Cliff, D. (2011).
\newblock Human-agent auction interactions: Adaptive-aggressive agents
  dominate.
\newblock In {\em Proceedings of the Twenty-Second International Joint
  Conference on Artificial Intelligence}.

\bibitem[Deffuant, 2006]{deffuant2006}
Deffuant, G. (2006).
\newblock {Comparing Extremism Propagation Patterns in Continuous Opinion
  Models.}
\newblock {\em Journal of Artificial Societies and Social Simulation}, 9(3):8.

\bibitem[Deffuant et~al., 2000]{deffuant2000}
Deffuant, G., Neau, D., and Amblard, F. (2000).
\newblock {Mixing Beliefs Among Interacting Agents.}
\newblock {\em Advances in Complex Systems}, 3:87--98.

\bibitem[Deffuant et~al., 2002]{deffuant2002}
Deffuant, G., Neau, D., and Amblard, F. (2002).
\newblock {How can extremism prevail? A study based on the relative agreement
  interaction model.}
\newblock {\em Journal of Artificial Societies and Social Simulation}, 5(4):1.

\bibitem[DeGroot, 1974]{deGroot}
DeGroot, M. (1974).
\newblock {Reaching a Consensus}.
\newblock {\em Journal of the American Statistical Association},
  69(345):118--121.

\bibitem[Duffy and Utku~{\"U}nver, 2006]{Duffy}
Duffy, J. and Utku~{\"U}nver, M. (2006).
\newblock {Asset Price Bubbles and Crashes with Near-Zero-Intelligence
  Traders}.
\newblock {\em Economic Theory}, 27:537--563.

\bibitem[Friedkin, 1999]{friedkin}
Friedkin, N. (1999).
\newblock {Choice Shift and Group Polarization}.
\newblock {\em American Sociological Review}, 64(6):856--875.

\bibitem[Gode and Sunder, 1993]{GodeandSunder}
Gode, D. and Sunder, S. (1993).
\newblock {Allocative Efficiency of Markets with Zero-Intelligence Traders:
  Market as a Partial Substitute for Individual Rationality}.
\newblock {\em Journal of Political Economy}, 101(1):119--137.

\bibitem[Hegselmann and Krause, 2002]{hegselmannandkrause}
Hegselmann, G. and Krause, U. (2002).
\newblock Opinion dynamics and bounded confidence: models, analysis and
  simulation.
\newblock {\em Journal of Artificial Societies and Social Simulationn}, 5(3):2.

\bibitem[{Hommes, C. and LeBaron, B.}, 2018]{hommes_lebaron_2018}
{Hommes, C. and LeBaron, B.}, editor (2018).
\newblock {\em Computational Economics: Heterogeneous Agent Modeling}.
\newblock North-Holland.

\bibitem[Krause, 2000]{krause}
Krause, U. (2000).
\newblock A discrete nonlinear and non-autonomous model of consensus formation.
\newblock {\em Proc.\ 4th Int.\ Conf.\ on Difference Equations}, pages 27--32.

\bibitem[Lomas, 2020]{myThesis}
Lomas, K. (2020).
\newblock Exploring narrative economics: A novel simulation platform that
  integrates automated traders with opinion dynamics.
\newblock Master's thesis, University of Bristol Department of Computer
  Science.

\bibitem[Meadows and Cliff, 2012]{reexaminingRA}
Meadows, M. and Cliff, D. (2012).
\newblock {Reexamining the Relative Agreement Model of Opinion Dynamics}.
\newblock {\em Journal of Artificial Societies and Social Simulation}, 15(4):4.

\bibitem[Meadows and Cliff, 2013]{RD}
Meadows, M. and Cliff, D. (2013).
\newblock {The Relative Disagreement Model of Opinion Dynamics: Where Do
  Extremists Come From?}
\newblock {\em 7th International Workshop on Self-Organizing Systems (IWSOS)},
  pages 66--77.

\bibitem[Shiller, 2017]{narrative_econ_paper}
Shiller, R. (2017).
\newblock {Narrative Economics}.
\newblock Technical Report 2069, Cowles Foundation, Yale University.

\bibitem[Shiller, 2019]{narrative_econ_book}
Shiller, R. (2019).
\newblock {\em {Narrative Economics: How Stories Go Viral \& Drive Major
  Economic Events}}.
\newblock {Princeton University Press}.

\bibitem[Smith, 1759]{theInvisibleHand}
Smith, A. (1759).
\newblock {\em {The Theory of Moral Sentiments}}.
\newblock Penguin Classics.

\bibitem[Smith, 1962]{VernonSmith}
Smith, V. (1962).
\newblock {An Experimental Study of Competitive Market Behaviour}.
\newblock {\em Journal of Political Economy}, 70(2):111--137.

\bibitem[Steindl et~al., 2015]{reactance}
Steindl, C., Jonas, E., Sittenhaler, S., Traut-Mattausch, E., and Greenberg, J.
  (2015).
\newblock {Understanding Psychological Reactance}.
\newblock {\em Zeitschrift f{\"u}r Psychologie}, 223(4):205--214.

\bibitem[Vytelingum, 2006]{vytelingum2006}
Vytelingum, P. (2006).
\newblock {\em The Stucture and Behavior of the Continuous Double Auction}.
\newblock PhD thesis, University of Southampton.

\end{thebibliography}

\end{document}